\documentclass[twocolumn]{aastex62}

\usepackage[version=4]{mhchem}

\newcommand\konkoly{Konkoly Observatory, Research Centre for Astronomy and Earth Sciences, E\"otv\"os Lor\'and Research Network (ELKH), Konkoly-Thege Mikl\'os \'ut 15-17, 1121 Budapest, Hungary}
\newcommand\csfk{CSFK, MTA Centre of Excellence, Konkoly-Thege Mikl\'os \'ut 15-17, 1121 Budapest, Hungary}
\newcommand\elte{ELTE E\"otv\"os Lor\'and University, Institute of Physics, P\'azm\'any P\'eter s\'et\'any 1/A, 1117 Budapest, Hungary}
\newcommand\mpia{Max Planck Institute for Astronomy, K\"onigstuhl 17, 69117 Heidelberg, Germany}
\received{December 16, 2022}
\accepted{January 19, 2023}
\submitjournal{ApJL}

\shorttitle{JWST/MIRI Spectroscopy of EX Lup}
\shortauthors{K\'osp\'al et al.}

\begin{document}

\title{JWST/MIRI Spectroscopy of the Disk of the Young Eruptive Star EX Lup in Quiescence}

\author[0000-0001-7157-6275]{\'Agnes K\'osp\'al}
\affiliation{\konkoly{}}
\affiliation{\csfk{}}
\affiliation{\mpia{}}
\affiliation{\elte{}}

\author[0000-0001-6015-646X]{P\'eter \'Abrah\'am}
\affiliation{\konkoly{}}
\affiliation{\csfk{}}
\affiliation{\elte{}}

\author{Lindsey Diehl}
\author[0000-0003-4335-0900]{Andrea Banzatti}
\affiliation{Department of Physics, Texas State University, 749 N Comanche Street, San Marcos, TX 78666, USA}

\author[0000-0003-4757-2500]{Jeroen Bouwman}
\affiliation{\mpia{}}

\author[0000-0003-2835-1729]{Lei Chen}
\affiliation{\konkoly{}}
\affiliation{\csfk{}}

\author[0000-0002-4283-2185]{Fernando Cruz-S\'aenz de Miera}
\affiliation{\konkoly{}}
\affiliation{\csfk{}}

\author[0000-0003-1665-5709]{Joel D. Green}
\affiliation{The University of Texas at Austin, Department of Astronomy, 2515 Speedway, Stop C1400, Austin, TX 78712, USA}
\affiliation{Space Telescope Science Institute, 3700 San Martin Drive, Baltimore, MD 02138, USA}

\author[0000-0002-1493-300X]{Thomas Henning}
\affiliation{\mpia{}}

\author[0000-0003-1817-6576]{Christian Rab}
\affiliation{University Observatory, Faculty of Physics, Ludwig-Maximilians-Universit\"at M\"unchen, Scheinerstr. 1, 81679, Munich, Germany}
\affiliation{Max-Planck-Institut f\"ur Extraterrestrische Physik, Giessenbachstr. 1, 85748, Garching, Germany}


\begin{abstract}
EX\,Lup is a low-mass pre-main sequence star that occasionally shows accretion-related outbursts. Here, we present JWST/MIRI medium resolution spectroscopy obtained for EX\,Lup fourteen years after its powerful outburst. EX\,Lup is now in quiescence and displays a Class II spectrum. We detect a forest of emission lines from molecules previously identified in infrared spectra of classical T Tauri disks: \ce{H2O}, OH, \ce{H2}, HCN, \ce{C2H2}, and \ce{CO2}. The detection of organic molecules demonstrates that they are back after disappearing during the large outburst. Spectral lines from water and OH are for the first time de-blended and will provide a much improved characterization of their distribution and density in the inner disk. The spectrum also shows broad emission bands from warm, sub-micron size amorphous silicate grains at 10 and 18$\,\mu$m. During the outburst, in 2008, crystalline forsterite grains were annealed in the inner disk within 1 au, but their spectral signatures in the 10 $\mu$m silicate band later disappeared. With JWST we re-discovered these crystals via their 19.0, 20.0, and 23.5$\,\mu$m emission, whose strength implies that the particles are at $\sim$3\,au from the star. This suggests that crystalline grains formed in 2008 were transported outwards and now approach the water snowline, where they may be incorporated into planetesimals. Containing several key tracers of planetesimal and planet formation, EX\,Lup is an ideal laboratory to study the effects of variable luminosity on the planet-forming material and may provide explanation for the observed high crystalline fraction in solar system comets.
\end{abstract}

\keywords{Protoplanetary disks (1300) --- Eruptive variable stars (476) --- Low mass stars (2050) --- Infrared spectroscopy (2285) --- stars: individual (EX Lup)}


\section{Introduction}
\label{sec:intro}

Circumstellar disks provide the material that forms stars and planetary systems. Dust grains in the disk make rocky planets and the cores of gas giants, while the forming planets may accrete their atmospheres directly from the disk's gas content or when the grains' ice mantles sublimate. Therefore, studying the composition of protoplanetary disks is needed to understand the initial mineralogical and chemical composition that the young planets inherit \citep[e.g.,][]{jorgensen2020,eistrup2022,wordsworth2022}.

Infrared spectroscopy is an excellent tool to investigate the warm material in protoplanetary disks. The small, amorphous silicate grains that the disks of \mbox{T Tauri} stars inherit from the interstellar medium have prominent broad features around $9.7\,\mu$m due to the Si–O stretching modes and around $18\,\mu$m due to O–Si–O bending modes \citep{henning2010}. During their evolution, these silicates may undergo thermal processing and form crystalline silicates such as forsterite and enstatite, which have multiple narrow features at infrared wavelengths \citep[e.g.][]{fabian2001}. The infrared regime is rich in rotational and rovibrational lines of gas-phase molecules and broad bands of molecular ices. These have been extensively studied using space missions like the Infrared Space Observatory \citep[e.g.,][and references therein]{vandishoeck2004}, the Spitzer Space Telescope \citep[e.g.,][]{boogert2008,pontoppidan2010}, the Herschel Space Observatory \citep[e.g.,][]{fedele2013,dent2013}, and now the JWST.

While circumstellar disks typically evolve on $10^5 - 10^6$\,yr time scales \citep{williams2011}, variable heating and changing irradiation during accretion bursts and outbursts may change them significantly faster \citep{fischer2022}. Due to the released energy, outbursts can have significant effects on the disk material, modifying its physical, chemical, and mineralogical properties \citep{molyarova2018, abraham2019}. As eruptions overlap with the earliest stages of planet formation, episodic accretion may have far-reaching consequences for disk evolution and planet formation \citep[e.g.,][]{cieza2016}.

The best studied example for accretion outburst induced changes is EX\,Lup, the prototype of the EXor class of young eruptive stars \citep{herbig1989}. In 2008, EX\,Lup exhibited its largest outburst ever observed ($\Delta$V$\sim$5\,mag). A multi-wavelength, multi-epoch investigation of this event revealed the annealing of amorphous silicates to crystalline grains, detected via their 10 and 11.3$\,\mu$m emission \citep{abraham2009}. In parallel, \citet{banzatti2012, banzatti2015} reported increased strength for \ce{OH} and \ce{H2O} lines during the outburst, while organics, such as \ce{C2H2} and \ce{HCN}, disappeared, possibly due to UV photodissociation. 

In the course of the year following the outburst, the spectral signatures of the crystalline grains at 10 and 11.3$\,\mu$m gradually weakened and later disappeared \citep{abraham2019}. During the 2008 outburst, cold forsterite emission also appeared between 15--35$\,\mu$m, which became even stronger after the end of the outburst in late 2008 - early 2009. As these long wavelength features were not present before the outburst, \citet{juhasz2012} proposed that the crystals formed in the inner disk and were transported to outer disk regions driven by a low velocity wind. Based on this idea, \citet{abraham2019} performed radiative transfer modeling of multi-epoch mid-infrared spectra. They predicted that the crystalline grains cooled down and are now located at $\geq$3\,au from the central star, but could be as far as $\sim$10\,au (with an expansion velocity of $\sim$3\,km\,s$^{-1}$). If the forsterite grains can reach the water snowline at 3.5\,au on the disk surface \citep{abraham2009}, they may bring crystalline material to newly forming planetesimals and reveal the details of the radial transport processes in the disk. After Spitzer but before JWST, we lacked the sensitivity and wavelength coverage to detect these cold crystals.

Apart from smaller bursts in 2011 and early 2022 \citep{abraham2019, zhou2022, kospal2022}, EX\,Lup has been in quiescence since 2008. While it may currently be similar to a typical, quiescent \mbox{T Tauri} disk, it also offers an opportunity to study possible residual effects from the 2008 large outburst. To survey the mid-infrared spectral features in EX\,Lup's disk, we observed it with the medium-resolution spectrometer (MRS) of JWST's Mid-Infrared Instrument (MIRI). Here, we present our first results based on this spectrum.


\begin{figure*}
\includegraphics[angle=0,width=\textwidth]{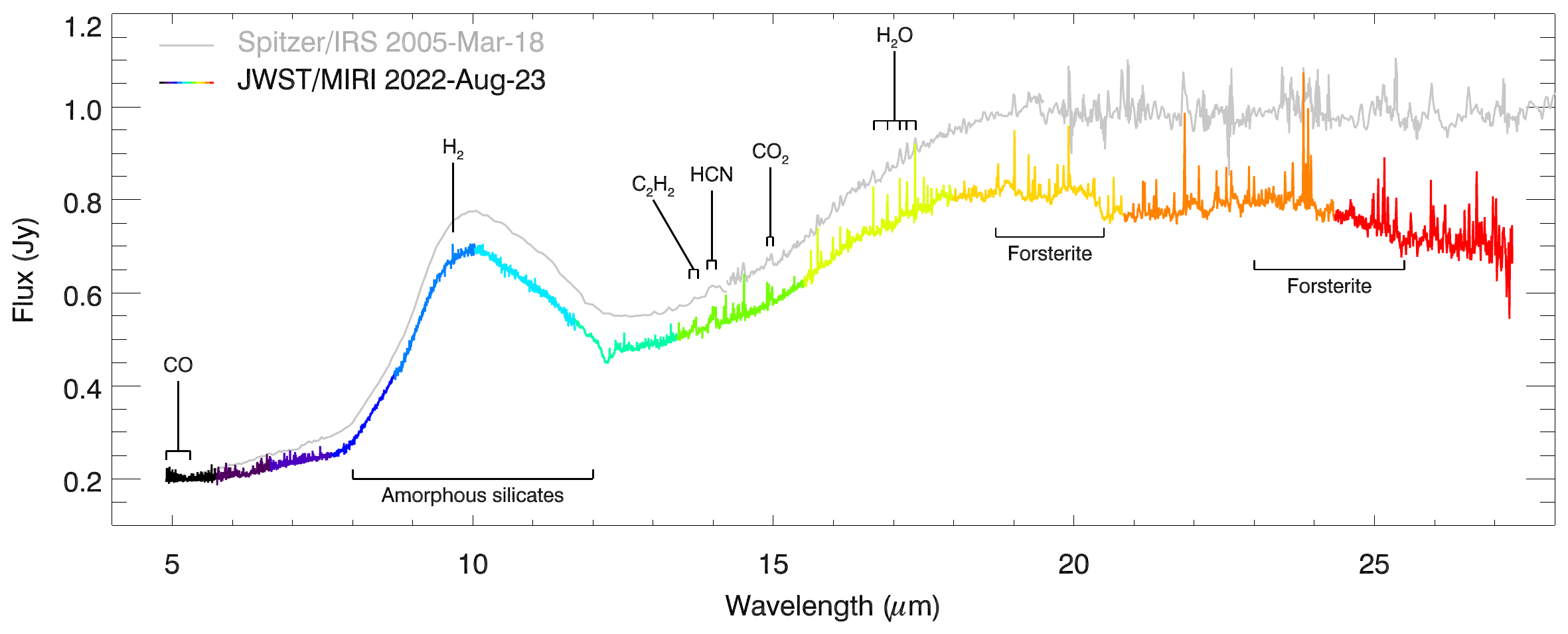}
\caption{JWST/MIRI MRS spectrum of EX\,Lup (color), shown together with an earlier Spitzer/IRS spectrum (gray), both taken in quiescence. The different colors mark the individual sub-bands of the JWST spectrum. We indicated some of the observed molecular and solid state features. We note that the spectrum contains hundreds of \ce{H2O} lines; we only marked a few of them as an illustration. The JWST spectrum shown here is available as Data behind the Figure.\label{fig:test_aper}}
\end{figure*}

\section{Observations and data reduction}
\label{sec:obs}

We obtained JWST/MIRI MRS spectroscopy of EX\,Lup on 2022 August 23 (GO 2209, PI: P.~\'Abrah\'am). We used all four channels and all three sub-bands, covering 
4.9--28$\,\mu$m at a spectral resolution of 3500--1500. We observed a separate background $193\farcs{}5$ away from EX\,Lup, centered at RA$_{\rm J2000}=16^{\rm h}03^{\rm m}12\fs{}63$, DEC$_{\rm J2000}=-40^{\circ}15'30\farcs{}42$. Exposure time was 344.1\,s for the science target and 41.6\,s for the background. We used the FASTR1 readout mode and 4-point dithering for EX\,Lup and no dithering for the background. We reduced the spectra by running the JWST pipeline version 1.8.2. Dedicated background exposures were subtracted on the exposure level to remove the thermal background. We further applied a residual fringe correction based on the empirical sine model fitting method described in, e.g., \citet{Kester03} and Kavanagh et al.~(in prep.). The correction fits and removes fringe residuals that are assumed to be low in amplitude after application of the fringe flat correction in the pipeline.

The spacecraft pointing information was about 1$\farcs$2 offset from the actual pointing inferred from both the MIRI integral field data and MIRI simultaneous images. Due to this, EX\,Lup was not well centered in the MRS field of view; a significant part of the source was cut off in two of the four dither positions observed in the short wavelength channels, where the field of view was the smallest. We discarded the affected dithers when constructing the spectral cubes for channels 1 and 2.

To check whether JWST spatially resolved the mid-infrared emission of EX\,Lup, we compared the 3D data cubes with similar measurements of a point source template. We used 10\,Lac, observed within the calibration program 1524 (PI: D.R.~Law) with the purpose of characterizing the point spread function of MIRI MRS. After rotating, shifting, scaling, and subtracting the 10\,Lac measurements from the EX\,Lup measurements channel by channel, we found that the residuals were typically below a few \%. A comparison between 2D Gaussian sizes fitted to EX\,Lup and 10\,Lac did not reveal any significant difference in the $4.9-26.5\,\mu$m range. At longer wavelengths, 10\,Lac was too faint to be detected reliably. Thus, EX\,Lup remained unresolved by MIRI.

We calculated the centroid for each image in the 3D data cube and used aperture photometry to extract the 1D spectrum using the default parameters for point sources. Flux levels of different channels and sub-bands (for the usable dither positions) were in general consistent within 5\% in the overlapping wavelength ranges. One exception is the channel 4 medium sub-band (20.8--24.3$\,\mu$m), which was lower by 10\%. We corrected these mismatches by applying multiplicative factors determined in the overlapping regions. No scaling was applied to channel 2 (7.6--11.7$\,\mu$m). The observed shifts between the spectral channels are consistent with the quoted 10\% absolute flux calibration uncertainty determined during the commissioning of the MRS instrument.\footnote{\href{https://www.stsci.edu/files/live/sites/www/files/home/jwst/documentation/_documents/jwst-science-performance-report.pdf}{https://www.stsci.edu/files/live/sites/www/files/home/jwst/\\documentation/\_documents/jwst-science-performance-report.pdf}} To the best of our knowledge, the only remaining instrumental artifact is the dip around 12.25$\,\mu$m due to a spectral leak.

To check the accuracy of the wavelength calibration of our spectrum, we measured the wavelengths of 139 strong non-blended water emission lines (\autoref{sec:res}) and computed the shifts compared to the theoretical wavelengths. According to the documentation, the wavelength calibration of MIRI MRS is accurate to within 1 spectral resolution element (85--200\,km\,s$^{-1}$ depending on channel and sub-band). We found that the measured shifts are less than 0.0015$\,\mu$m below 10$\,\mu$m and less than 0.003$\,\mu$m above 10$\,\mu$m, much better than the nominal accuracy.

The final spectrum is plotted in \autoref{fig:test_aper}. As representative values, we reached signal-to-noise ratios (S/N) of 120 at 9.0$\,\mu$m, 100 at 15.5$\,\mu$m, and 60 at 25.0$\,\mu$m on the continuum. 


\section{Results and analyses}
\label{sec:res}

\begin{figure*}
\includegraphics[angle=0,width=\textwidth]{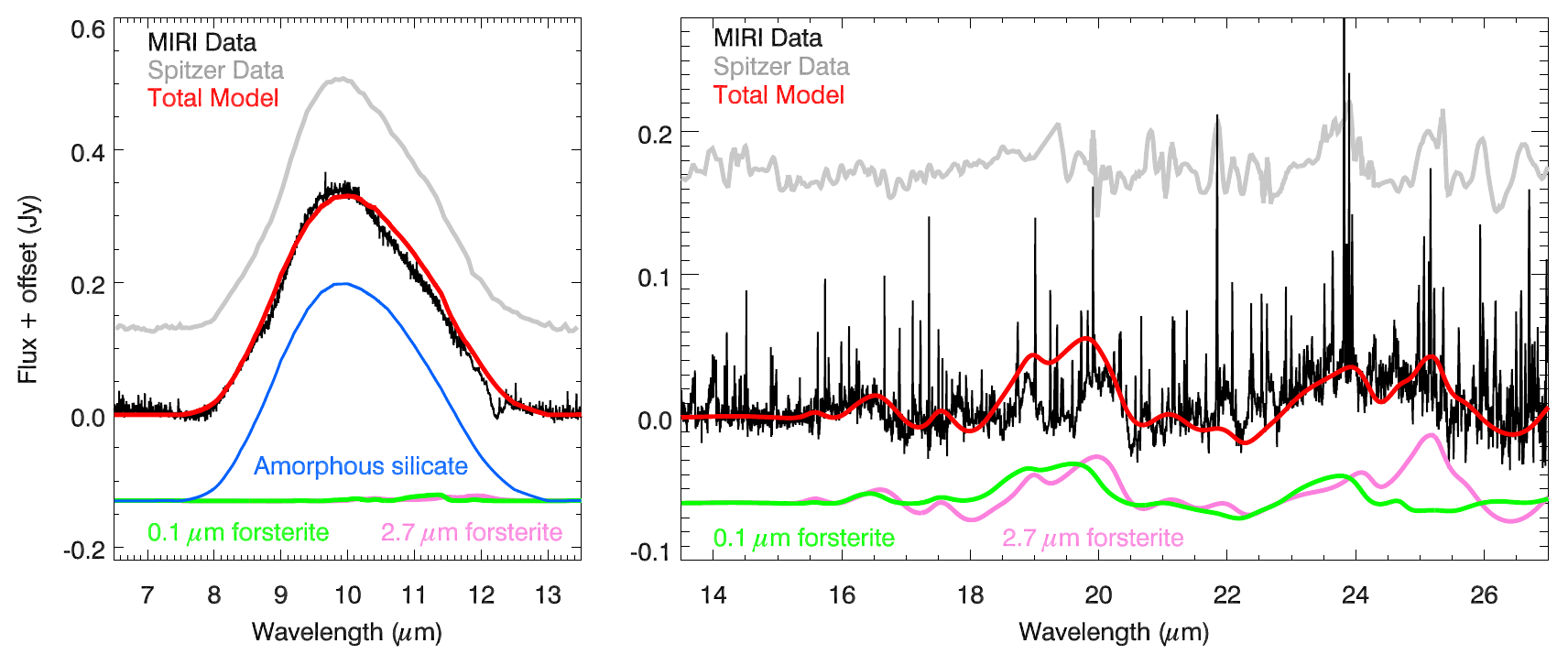}
\caption{Continuum-subtracted JWST/MIRI spectrum of EX\,Lup with various dust models, and with the Spitzer/IRS quiescent spectrum from 2005. The model curves were shifted for clarity.\label{fig:dust}}
\end{figure*}

\subsection{Spectral Energy Distribution}
 
The MIRI MRS spectrum of EX\,Lup, plotted in $F_{\nu}$, slowly rises towards longer wavelengths between 4.9--8$\,\mu$m, after which a prominent 10$\,\mu$m silicate emission feature is visible. Above 13$\,\mu$m, the spectrum increases again until $\sim$18$\,\mu$m, where a shallow, broad 18$\,\mu$m silicate feature begins. Apart from some weaker, narrower silicate features between 18--24$\,\mu$m the spectrum is approximately flat here, then slowly decreases toward longer wavelengths.

At the date of the MIRI observation (2022 August), EX\,Lup was in quiescence following a medium-size burst occurring between 2022 February--April (Cruz-S\'aenz de Miera et al.~in prep.). Thus, we compared the spectral shape with an earlier quiescent spectrum obtained with Spitzer/IRS in 2005 March (\autoref{fig:test_aper}; \citealt{sipos2009,abraham2009}). The two spectra have similar shapes with slight differences. The flux levels are the same at 5$\,\mu$m, but the JWST spectrum has a shallower slope at 5--8$\,\mu$m. Between 8--18$\,\mu$m, the two spectra have practically identical shapes, but JWST measured 12\% lower flux than Spitzer. Above 18$\,\mu$m, the two spectra again have slightly different slopes: the earlier one is flat, the newer one decreases towards longer wavelengths. We also compared the JWST spectrum with the average of four ISOPHOT/PHOT-S spectra obtained in 1997 February-September \citep{kospal2012} and found that in the overlapping wavelength range (5.8--11.6$\,\mu$m) they are consistent with MIRI within the uncertainties. These results demonstrate that the MIRI spectrum represents well the quiescent state of EX\,Lup.

\subsection{Dust Features}
\label{sec:dust}

The characteristic strong, triangular shape of the 10$\,\mu$m silicate emission feature in the 2005 Spitzer spectrum (\autoref{fig:dust}) was interpreted by \citet{abraham2009} as the signature of amorphous sub-micron-sized grains. \citet{sipos2009} fitted this profile and found that the mass fraction of crystalline silicates was $<$2\%. Since the 10$\,\mu$m feature in our MIRI spectrum has a very similar shape to the Spitzer/IRS, we conclude that the disk of EX\,Lup contained only small amounts of warm forsterite or enstatite in 2022 August. 

In \autoref{fig:dust} we compared our JWST spectrum with the 10$\,\mu$m emission profile of amorphous dust grains using the opacity curves of olivine and pyroxene type silicates. We derived dust opacities using the OPACITYTOOL software \citep{toon1981, min2005, woitke2016}, which is based on the distribution of hollow spheres (DHS) theory (Min et al. 2005). We took the complex refractive index for amorphous silicate from the literature \citep{dorschner1995}. Following \citet{sipos2009}, we used 0.1$\,\mu$m-sized olivines and 1.5$\,\mu$m-sized pyroxenes with a ratio of 2:1. After continuum subtraction, the measured spectral shape can be well reproduced by adopting $T\approx1200$\,K for the underlying blackbody emission, but higher temperatures up to 1500\,K give similarly good results. 

At longer wavelengths the MIRI spectrum exhibits several narrower emission peaks at 16.3, 19.0, 20.0, and 23.5$\,\mu$m. To check if they originate from crystalline silicates, we compared our spectrum with opacity curves of forsterite grains \citep{servoin1973}. In \autoref{fig:dust} we plotted the infrared spectrum of small 0.1$\,\mu$m (green) and larger 2.7$\,\mu$m (pink) forsterite crystalline grains. We multiplied the opacities by a blackbody curve of 150\,K to represent a cold dust population. We found that the small forsterite grains could be responsible for the features peaking around 16.3, 19.0, and 23.5$\,\mu$m. The additional presence of larger forsterite grains may explain the feature at 20.0$\,\mu$m. 

As our model in \autoref{fig:dust} shows, small forsterite grains also have an emission peak at 11.3\,$\mu$m, whose strength depends strongly on the temperature. This feature is not detected in the observed JWST spectrum. Tests with various dust temperatures show that it would be visible for $\geq$175\,K. In conclusion, the dust composition in the EX\,Lup disk is dominated by warm (T$\geq$1200\,K) amorphous grains, but a colder (T$\leq$150\,K) crystalline silicate, probably forsterite, component is also present. 

\subsection{Gas features}

\begin{figure*} 
\includegraphics[angle=0,width=\textwidth]{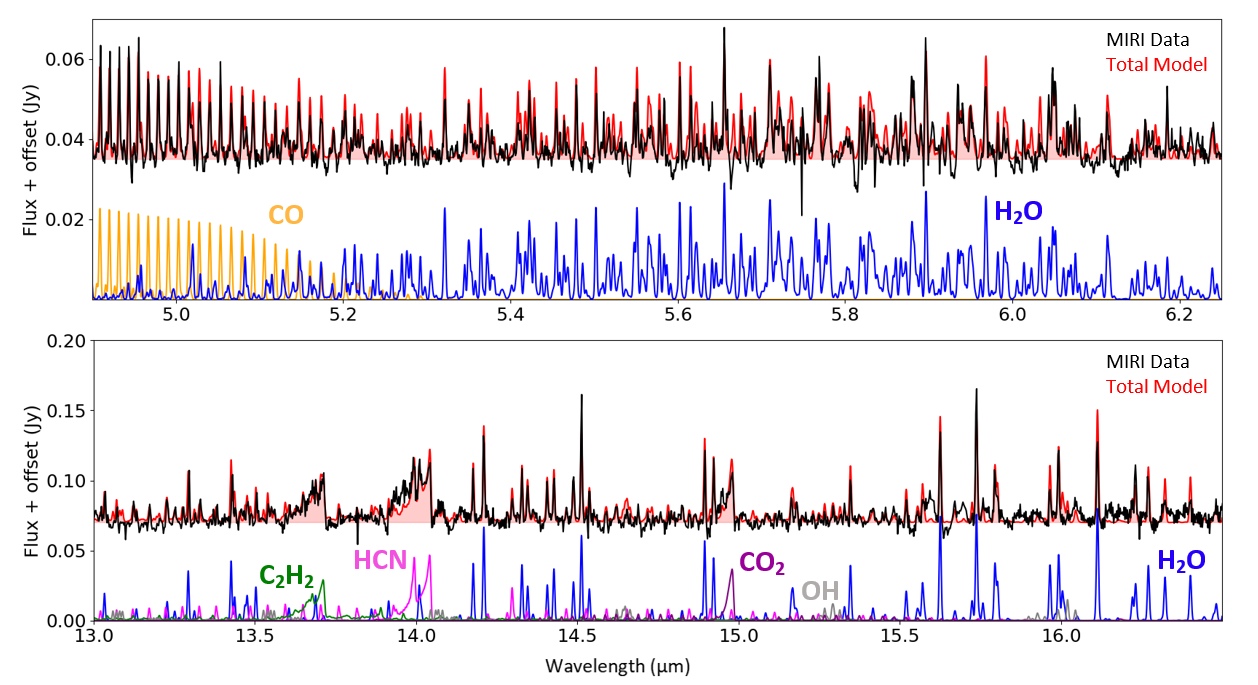}
\caption{Continuum-subtracted MIRI spectrum with slab models for the identification of molecular lines.\label{fig:slab_models}}
\end{figure*}

The MIRI spectrum of EX\,Lup also includes a forest of gas emission lines from molecules previously observed in \mbox{T Tauri} disks \citep{cn08,cn11,bitner2008,lahuis2007}: \ce{H2O}, OH, \ce{H2}, HCN, \ce{C2H2}, and \ce{CO2}. These molecules were identified in EX\,Lup either during its quiescent phase in 2005 or during outburst in 2008 using Spitzer/IRS. Line strengths strongly changed between the two epochs, with organics disappearing from the spectrum and stronger \ce{OH} lines, suggesting effects from UV photodissociation \citep{banzatti2012}. Additionally, \ce{CO} ro-vibrational lines in the fundamental band were previously observed in EX\,Lup during outburst in 2008 \citep{goto2011} and during quiescence in 2014 \citep{banzatti2015} using VLT/CRIRES. These data suggested the presence of hot molecular gas in the inner disk during outburst.

The new MIRI data show the 4.9--27$\,\mu$m mid-infrared spectrum of EX\,Lup for the first time since 2008, covering all of the above mentioned molecular lines at high resolution, allowing us to study longer-term chemical changes induced by an accretion outburst. In this first paper, we only provide a brief overview of the most prominent lines observed in the spectrum and illustrate the approximate match to slab models. The in-depth analysis of the molecular emission is left to an upcoming paper.

For line identification, we produced synthetic spectra using a slab model in local thermodynamic equilibrium following \citet{banzatti2012}, with molecular data from the latest HITRAN release \citep{gordon2022}. We adopted parameters from previous works and adjusted them by eye to match the relative strength of gas emission lines in the MIRI data. \autoref{fig:slab_models} displays portions of the continuum-subtracted spectrum showing all the major molecules observed. Hundreds of H$_2$O lines are detected from ro-vibrational bands below 8$\,\mu$m and from pure rotational transitions between 12--28~$\mu$m. Compared to previous Spitzer spectra, most of these lines are now de-blended in the MIRI spectrum, with some blending still visible at the shortest wavelengths. In \autoref{fig:slab_models}, we show two slab models for water using properties previously found from fits to high-resolution ground-based spectra of other \mbox{T Tauri} disks \citep{banz23}: $T \sim 1000\,$K for the ro-vibrational lines below 6.6~$\mu$m and $T \sim 600\,$K for the rotational lines at 13--16~$\mu$m, and a similar column density of $\approx 10^{17}$--$10^{18}$~cm$^{-2}$. 

For \ce{CO} we adopted properties found previously in high-resolution spectra that covered lines up to $J=50$ \citep{banz22}: $T \sim 800\,$K and $N_{\rm CO} \sim 10^{19}\,$cm$^{-2}$. Since MIRI has insufficient resolution to distinguish the broad (BC) and narrow (NC) velocity components observed in ground-based data, we adopted here $T$ and $N$ in between those estimated for BC and NC in other disks \citep[see Figure 11 in][]{banz22}. The relative weakness of the $v=2-1$ lines suggests sub-thermal vibrational excitation, so we used a lower vibrational temperature $T_{\rm vib} \sim 450\,$K.

Detection of ro-vibrational emission from \ce{C2H2}, \ce{HCN}, and \ce{CO2} demonstrates that the organic molecules are back after disappearing during the 2008 outburst. The Q-branch emission lines (near 13.7~$\mu$m, 14.0~$\mu$m, and 14.9~$\mu$m respectively) are still blended at the resolution of MIRI, but several weak lines are now detected for the first time from the P and R branches. For \ce{HCN}, multiple peaks in the Q branch are distinguished with MIRI. The slab models in \autoref{fig:slab_models} assumed $T \sim 300-800$~K and column density of $\approx 10^{15}$--$10^{16}$~cm$^{-2}$ as found from previous analyses of Spitzer spectra of \mbox{T Tauri} disks \citep{cn11,salyk11}.

Multiple weak lines from \ce{OH} are detected throughout the spectrum. Mid-infrared \ce{OH} emission has previously shown very high temperatures indicative of production from UV photodissociation of water \citep{cn14}, and was found to increase during the 2008 outburst in EX\,Lup \citep{banzatti2012}. In \autoref{fig:slab_models}, we use $T \sim 4000$~K and $N \sim 10^{16}$~cm$^{-2}$.

Several rotational lines of molecular hydrogen (\ce{H2}) are detected from 0--0 S(8) at 5.05$\,\mu$m to 0--0 S(1) at 17.03$\,\mu$m \citep{roueff2019}. Some of the lines are blended with \ce{CO} or \ce{H2O}, but S(1), S(3), S(4), S(5), and S(7) are well resolved.

Lines of longer chain hydrocarbons (C$_2$H$_4$ at 10.5$\,\mu$m) or aromatic hydrocarbons (benzene, C$_6$H$_6$, at 5.1 and 5.5$\,\mu$m) were not found. We also searched for \ce{CH3OH} (9.7$\,\mu$m), \ce{CH4} (7.6$\,\mu$m), \ce{NH3} (10.1--10.9$\,\mu$m), and \ce{H2CO} (5.7$\,\mu$m), but found no such lines in the MIRI spectrum of EX\,Lup at this stage of the data reduction and analysis.


\section{Discussion}
\label{sec:conc}

\subsection{Quiescent Disk Variability}

In quiescence, EXors have accretion rates similar to regular \mbox{T Tauri} stars \citep{lorenzetti2012a}, therefore their disks should also be similar to those of normal \mbox{T Tauri} disks. Indeed, the MIRI spectrum of EX\,Lup is similar to typical \mbox{T Tauri} disk spectra \citep[e.g., ][]{furlan2006,kesslersilacci2006}. \citet{sipos2009} studied the quiescent spectral energy distribution (SED) of EX\,Lup and found that its shape is very similar to the median SED of \mbox{T Tauri} stars in the Taurus star-forming region. Interestingly, their radiative transfer modeling revealed that the dust disk of EX\,Lup does not extend inward until the dust sublimation radius (0.06\,au), but it has a larger inner radius at about 0.2--0.3\,au.

While the general shape of the 5--27$\,\mu$m spectrum of EX\,Lup is very similar between 2005 (Spitzer) and 2022 (JWST), the object was somewhat ($<$25\%) fainter in 2022. Such small infrared variability is not unusual for EX\,Lup in quiescence, because multiple quiescent Spitzer spectra (2004 August, 2005 March, and 2009 April) also show differences up to 20\%. A comparison of Spitzer and ISO spectra of EX\,Lup by \citet{kospal2012} also revealed only minor changes.

Interestingly, the slope of the SED in the 5--8$\,\mu$m and 24--27$\,\mu$m ranges as measured by JWST is different from those in 2005. Such variations of the infrared spectral slope, and in certain cases, anti-correlation between shorter and longer mid-infrared wavelengths with a pivot point in-between was noticed in other young stars and has been explained with variable inner rim height and variable shadowing of the outer disk \citep{muzerolle2009, flaherty2011, espaillat2011, kospal2012}. In EX\,Lup, the shallower slope at shorter wavelengths may indicate more warm material or a larger scale height around the inner edge of the dust disk. The radius of the inner edge might also have changed between the two epochs. The decreasing slope at longer wavelengths as opposed to the earlier flat SED may point to the possibility that the outer part of the disk received less illumination in 2022 than in 2005. 

\subsection{Silicate Crystals Towards the Snowline}

\begin{figure*}
\begin{center}
\includegraphics[angle=0,height=0.3\textheight]{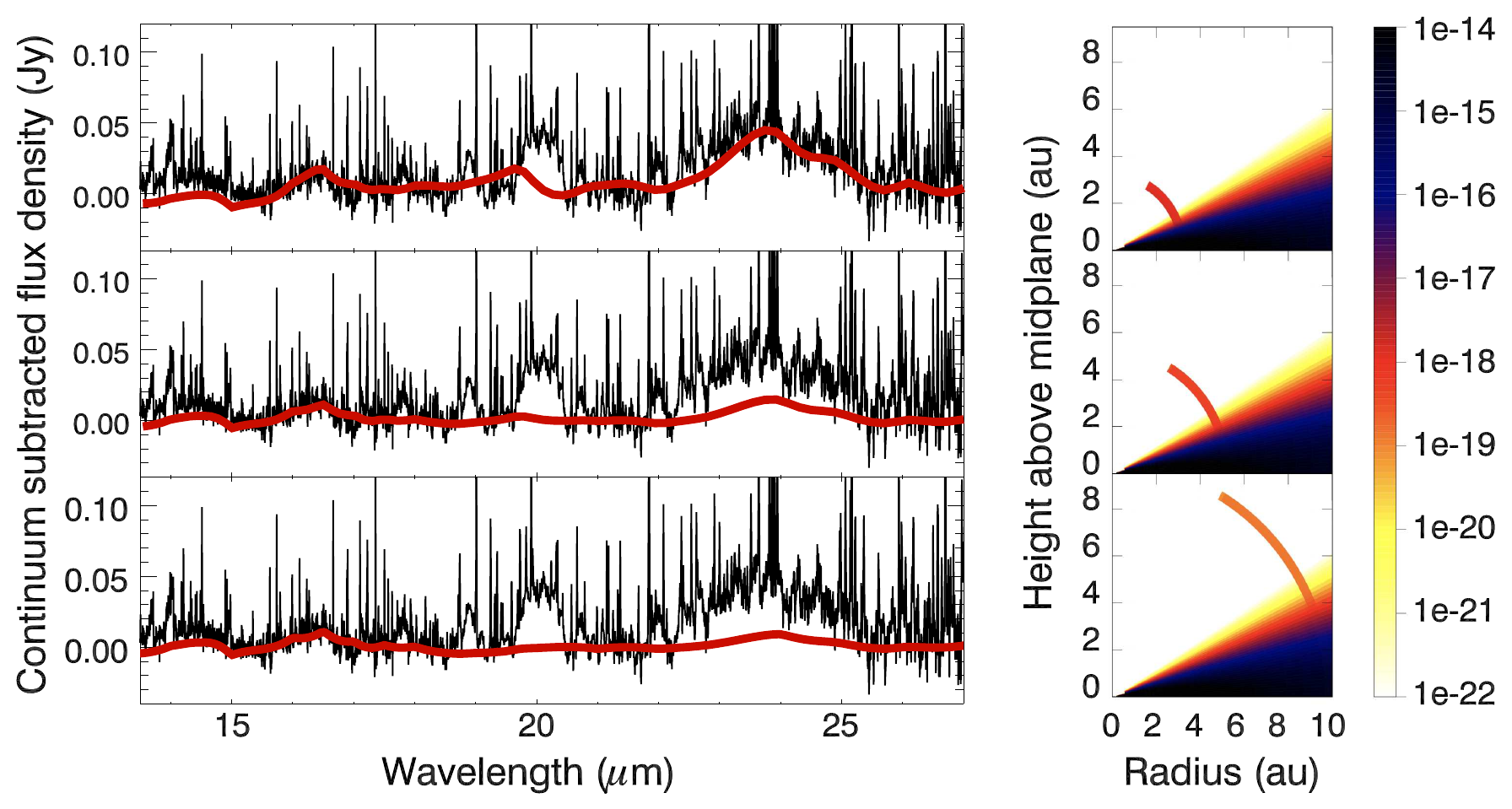} 
  \caption{Left: continuum subtracted JWST/MIRI spectrum of EX\,Lup (black) and model spectra including an amorphous disk and a hollow crystalline sphere placed at different radii (red). 
  Right: density distributions of the various models (in g\,cm$^{-3}$). Models are from fig.~2 of \citet{abraham2019}.}
\label{fig:spectra_JWST}
\end{center}
\end{figure*}

In \autoref{sec:dust} we showed that the profile of the 10\,$\mu$m silicate feature in our JWST spectrum is very similar to that observed with Spitzer/IRS in 2005, prior to the large outburst. Thus, JWST confirmed that the crystalline forsterite grains, produced in 2008 through thermal annealing on the disk surface between 0.3 and 0.7\,au and detected via their 10 and 11.3$\,\mu$m emission by \citet{abraham2009,abraham2019}, are not present in the inner disk any more. This result is consistent with VLTI/MIDI data from 2013 and VLT/VISIR data from 2016, which also failed to detect the crystalline features at 10 and 11.3$\,\mu$m, although at lower significance level \citep{abraham2019}.

The hypothesis of \citet{juhasz2012} and \citet{abraham2019}, that the forsterite grains formed in 2008 still exist but were transported to outer colder disk regions (\autoref{sec:intro}), could not be verified after the end of the cryogenic mission of Spitzer (May 2009) until now. The unambiguous detection of cold crystalline features in the 16--26$\,\mu$m range of the JWST/MIRI spectrum, however, implies that we very likely re-discovered the forsterite grains that were inaccessible for the available instrumentation for more than a decade.

To infer the radial location of the cold crystals, we compared the measured crystalline profiles with synthetic spectra computed by \citet{abraham2019} for a model containing a disk made of amorphous dust grains and an expanding hollow sphere made of crystalline forsterite, assuming three different radii for the sphere: 3\,au, 5\,au, and 9.68\,au. \autoref{fig:spectra_JWST} presents this comparison. The strength of the 23.5\,$\mu$m band suggests that our JWST measurement is most consistent with a radial distance of $\sim$3\,au for the current location of the crystalline material.

The derived radial distance of $\sim$3\,au suggests that the crystalline grains formed in 2008 may now be in the vicinity of the water snowline: a disk region where the temperature drops to 150--170\,K and water molecules freeze out onto grain surfaces (at 3.5\,au on the surface of EX\,Lup's disk). The temperature of $\leq$150\,K, estimated for the crystalline dust in \autoref{sec:dust}, is fully consistent with this conclusion. We have very limited information on the spatial distribution of the crystals. Their geometry is probably determined by the properties of the wind that provides the momentum for the transportation of the grains.

\citet{abraham2019} proposed that the crystals were at 3\,au already in 2013. As the current JWST spectrum indicates that the crystalline grains are still at 3\,au, their radial expansion might have slowed down. Thus, they are not expected to move significantly further away from the star in the future. However, they may not stay undisturbed in their current location either. In 1955/56 EX\,Lup underwent a large outburst, similar to the one in 2008 \citep{herbig1977}, and it is reasonable to assume that a comparable amount of crystalline silicates was then produced. Those grains could have also moved outward, but the Spitzer spectrum in 2005 exhibited no cold forsterite emission (\autoref{fig:dust}), suggesting that something happened with the cooled-down crystalline grains on decadal timescales.

Vertical mixing in the disk could remove the crystals from the surface. Fig.~6 in the Supplementary Material of \citet{abraham2009} shows that at a radial distance of 3\,au, a turbulent viscosity parameter $\alpha$ of a few times $10^{-2}$ gives a vertical mixing timescale of $\leq$50\,yr. Alternatively, crystalline particles may start growing ice mantles as they move outward, leading to the non-detection of well-defined mid-infrared spectral features due to the increased grain size. Using Eqn.~12 from \citet{rab2017}, and adopting representative parameters for the EX\,Lup disk at 3\,au close to the vertical water ice line, we obtain a freeze-out timescale of about 3\,yr. Based on these considerations, we anticipate changes in the strength of the crystalline spectral features during the next years/decades in EX\,Lup. Future monitoring with JWST will be invaluable to follow this evolution.

When the crystalline grains reach the snowline, they may acquire ice mantles, and the increased collisional cross section and stickiness would facilitate future incorporation into planetesimals, including icy protocomets. If all the crystalline grains produced in the 2008 outburst end up in the water snowline region, that would add $\sim$1.9$\times$10$^{23}$\,g (equal to 10$^4$ times the mass of comet Hale–Bopp) to the crystalline dust reservoir \citep{abraham2019}. If similar large outbursts occur in every 50 years for an extended period of 10$^5$ years, the total yield would be 2$\times$10$^{-7}$\,M$_{\odot}$. Based on these numbers, we propose that large eruptions of young stars may be a significant source of crystalline grains in planetesimals and ultimately in the planet-building material.

\subsection{Future prospects}

The MIRI spectrum in 2022 provides the first opportunity to study mid-infrared molecular lines since the 2008 outburst. The organic molecules that disappeared in 2008 are now detected again. Modeling of these (and other) lines in the JWST spectrum and a detailed study of their time variability will be the topic of a future paper (Diehl et al.~in prep.). The increased spectral resolution of JWST/MIRI MRS (as compared to Spitzer/IRS) and the high S/N of the spectrum will allow us to de-blend and fit the various molecules, providing more stringent constraints than ever before on the thermal structure and molecular abundances in EX\,Lup's disk. We demonstrated that the spectrum also provides an unprecedentedly rich information on various silicate components and their temperature distribution, allowing the characterization of the planet and comet forming solids. Re-observing EX\,Lup with the JWST will provide a unique opportunity to monitor long-term chemical and mineralogical changes in the disk due to the variable accretion.


\vspace*{5mm}

\begin{acknowledgments}
We thank the reviewer for useful comments that helped to improve the manuscript. This project has received funding from the European Research Council (ERC) under the European Union's Horizon 2020 research and innovation programme under grant agreement No 716155 (SACCRED). CHR is grateful for support from the Max Planck Society and acknowledges funding by the Deutsche Forschungsgemeinschaft (DFG, German Research Foundation) - 325594231. This work is based on observations made with the NASA/ESA/CSA James Webb Space Telescope. The data were obtained from the Mikulski Archive for Space Telescopes (MAST) at the Space Telescope Science Institute, which is operated by the Association of Universities for Research in Astronomy, Inc., under NASA contract NAS 5-03127 for JWST. These observations are associated with programs \#2209 and \#1524. The specific observations analyzed can be accessed via \dataset[ http://doi.org/10.17909/7m42-er02]{ http://doi.org/10.17909/7m42-er02}.
\end{acknowledgments}

\facilities{JWST, Spitzer}


\bibliography{main}{}
\bibliographystyle{aasjournal}

\end{document}